\documentclass[conference]{IEEEtran}

\usepackage{booktabs}

\usepackage{amsmath}

\usepackage{subcaption}

\usepackage[utf8]{inputenc}
\usepackage[T1]{fontenc} 
\usepackage[english]{babel}
\usepackage{textcomp}

\usepackage{xspace}
\usepackage{comment}

\newcommand{\abstraction}{HeTM\xspace}
\newcommand{\system}{SHeTM\xspace} %Speculative Heterogeneous Transactional Memory System

\usepackage{amsmath,amssymb,amsfonts}
\usepackage{algorithm}
\usepackage{graphicx}
\usepackage{sidecap}
\usepackage{textcomp}
\def\BibTeX{{\rm B\kern-.05em{\sc i\kern-.025em b}\kern-.08em
		T\kern-.1667em\lower.7ex\hbox{E}\kern-.125emX}}

\usepackage{xspace}
\newcommand{\hetm}{HeTM\xspace}
\newcommand{\sys}{SHeTM\xspace}
\newcommand{\mem}{STMR\xspace}
\newcommand{\memShadow}{STMR$^S$\xspace}
\newcommand{\memWork}{STMR$^W$\xspace}

\graphicspath{{./resources/}}
\DeclareGraphicsExtensions{.pdf,.png,.jpg,.eps}
\usepackage{afterpage}
\usepackage{xcolor}

\usepackage[autostyle]{csquotes}

\usepackage{microtype}
\newcommand{\zerodisplayskips}{%
	\setlength{\abovedisplayskip}{0pt}%
	\setlength{\belowdisplayskip}{0pt}%
	\setlength{\abovedisplayshortskip}{0pt}%
	\setlength{\belowdisplayshortskip}{0pt}}
\appto{\normalsize}{\zerodisplayskips}
\appto{\small}{\zerodisplayskips}
\appto{\footnotesize}{\zerodisplayskips}

\usepackage{algorithmicx}
\usepackage[noend]{algpseudocode}

\makeatletter
\newlength{\algsectionrulewidth}

\makeatother

\usepackage{url}
\usepackage[bookmarks=false]{hyperref}
\hypersetup{final}

\newcommand\blfootnote[1]{%
	\begingroup
	\renewcommand\thefootnote{}\footnote{#1}%
	\addtocounter{footnote}{-1}%
	\endgroup
}

\begin{document}

\title{HeTM: Transactional Memory for \\Heterogeneous Systems}

% author names and affiliations (IEEE Format)
\author{

\IEEEauthorblockN{\vspace{8pt}\LARGE{Daniel Castro$^1$, Paolo Romano$^2$, Aleksandar Ilic$^3$, and Amin M. Khan$^4$}}
\IEEEauthorblockA{
$^{1,2,3}$ \textit{INESC-ID \& Instituto Superior T\'{e}cnico}, 
\textit{Universidade de Lisboa},
Lisbon, Portugal. \\
\vspace{6pt} $^4$ \textit{Department of Computer Science}, \textit{UiT The Arctic University of Norway}, Troms\o, Norway. \\
$^1$daniel.castro@tecnico.ulisboa.pt, $^2$romanop@gsd.inesc-id.pt, $^3$aleksandar.ilic@inesc-id.pt, $^4$amin.khan@uit.no\\
\vspace{-1.2cm}}

%\and

%\IEEEauthorblockN{Paolo Romano}
%\IEEEauthorblockA{
%\textit{INESC-ID. Instituto Superior T\'{e}cnico} \\
%\textit{Universidade de Lisboa}\\
%Lisbon, Portugal \\
%romanop@gsd.inesc-id.pt
%}

%\and

%\IEEEauthorblockN{Aleksandar Ilic}
%\IEEEauthorblockA{
%\textit{INESC-ID. Instituto Superior T\'{e}cnico} \\
%\textit{Universidade de Lisboa}\\
%Lisbon, Portugal \\
%aleksandar.ilic@tecnico.ulisboa.pt
%}

%\and

%\IEEEauthorblockN{Amin M. Khan}
%\IEEEauthorblockA{
%\textit{Department of Computer Science} \\
%\textit{UiT The Arctic University of Norway}\\
%Troms\o, Norway \\
%amin.khan@uit.no
%}

} % \author

\maketitle

%TODO: Skip page numbers for CR
\pagestyle{plain}

\begin{abstract}

Modern heterogeneous computing architectures, which couple multi-core CPUs with discrete many-core GPUs (or other specialized hardware accelerators), enable unprecedented peak performance and energy efficiency levels. However, developing applications that can take full advantage of the potential of heterogeneous systems is a notoriously hard task.
This work takes a step towards reducing the complexity of programming heterogeneous systems by introducing the abstraction of Heterogeneous Transactional Memory (\abstraction). \abstraction provides programmers with the illusion of a single memory region, shared among the CPUs and the (discrete) GPU(s) of a heterogeneous system, with support for atomic transactions. 
Besides introducing the abstract semantics and programming model of \abstraction, we present the design and evaluation of a concrete implementation of the proposed abstraction, referred herein as Speculative \abstraction (\system).
\system makes use of a novel design that leverages speculative techniques, which aims at hiding the inherently large communication latency between CPUs and discrete GPUs and at minimizing inter-device synchronization overhead.
%\system is based on a modular and extensible design that allows for easily integrating alternative TM implementations on the CPU's and GPU's sides, which allows the flexibility to adopt, on either side, the TM implementation (e.g., in hardware or software) that best fits the applications' workload and the architectural characteristics of the processing unit.
%
We demonstrate the efficiency of the \system via an extensive quantitative study based both on synthetic benchmarks and on a popular object caching system.
\end{abstract}


\iffalse % Only for ACM Templates
%
% The code below should be generated by the tool at
% http://dl.acm.org/ccs.cfm
% Please copy and paste the code instead of the example below.
%


\fi

\blfootnote{%\section*{Acknowledgement}

This work is funded by Fundação para a Ciência e a Tecnologia (FCT) OE component with references PTDC/EEIS-CR/1743/2014 and PTDC/EEI-COM/29271/2017 and by Programa
Operacional de financiamento POR Lisboa supported by FEDER with references Lisboa-01-0145-FEDER-031901 (PTDC/CCI-COM/31901/2017) and Lisboa-01-0145-FEDER-029271 (UID/CEC/50021/2019).
Amin's work is funded by the Research Council of Norway
under PREAPP project (grant n$^{\circ}$~231746/F20) and eX3 project (grant n$^{\circ}$~270053).
The authors would also like to thank the anonymous reviewers for their valuable comments and helpful suggestions.
}

\vspace{-0.5cm}
\section{Introduction}
\label{sec:intro}

Single-core performance of central processing units (CPUs) have reached a plateau in the last decade. In order to enable further increases of the processing capacity, while attaining high energy efficiency, modern computing architectures have henceforth adopted two key  paradigms, namely \textit{parallelism} and \textit{heterogeneity}. As a result, nowadays, heterogeneous  architectures that combine multi-core CPUs with many-core GPUs (or similar  co-processors, e.g., TPUs~\cite{Jouppi2016TPU}) have become the \textit{de facto} standard in a broad range of domains that include HPC, servers and mobile devices.

Unfortunately, though, developing applications that take full advantage of the raw performance potential of modern massively parallel, heterogeneous architectures is a notoriously hard task. This has fostered, over the last year, intense research efforts aimed at developing new abstractions and programming paradigms for reducing the complexity of software development for modern heterogeneous platforms.

This work focuses on one key problem that arises when developing concurrent applications, whose complexity is exacerbated when considering massively parallel heterogeneous architectures, namely how to regulate access to shared data in a scalable way.
% (HeTM, pronounced as hey-tee-em)
We tackle this problem by proposing the abstraction of Heterogeneous Transactional Memory (\abstraction). The \abstraction abstraction
% stands at the intersection of
combines two paradigms for concurrent programming, i.e., Transactional Memory (TM)~\cite{Herlihy93,Shavit1995STM} and Shared Memory (SM)~\cite{DSM1,DSM}, by providing the illusion of shared memory regions that are seamlessly accessed by CPUs and GPUs of a heterogeneous system, whose concurrent accesses can be synchronized via atomic transactions.

A large body of research has been devoted over the last years to investigate efficient implementations of both the TM and SM abstractions. 
%
%On the SM front, several implementations by major industrial players in the heterogeneous computing landscape, e.g., NVIDIA~\cite{Harris2013UM}.
% removed  and OpenAcc~\cite{Farber2016OpenACC}
Major industrial players in the heterogeneous computing landscape put a considerable amount of effort on implementing the SM abstraction, e.g., NVIDIA's Unified Memory~\cite{Harris2013UM}, which represent a strong evidence that the industrial world perceives the benefits, in terms of ease of programming. However, existing SM implementations for heterogeneous systems provide programmers with low-level synchronization primitives, such as atomic operations and locks, exposing programmers to another well-known source of complexity: the need of designing efficient, yet provably correct, techniques to regulate concurrent access to shared data.
%These efforts represent a strong evidence that the industrial world perceives the benefits, in terms of ease of programming, stemming from the adoption of the SM paradigm as compelling. However, existing SM implementations for heterogeneous systems provide programmers with low-level synchronization primitives, such as atomic operations and locks, exposing programmers to another well-known source of complexity: the need of designing efficient, yet provably correct, techniques to regulate concurrent access to shared data.

This is a notoriously hard problem, as designing efficient fine-grained locking protocols is a complex and error prone task~\cite{TM-easy} that can compromise one of the key principles of modern software development, i.e., software composability~\cite{composableTM}. TM addresses exactly this problem: thanks to the abstraction of atomic transactions, programmers only need to specify \textit{which} set of operations/code blocks have to be executed atomically, delegating to the TM implementation the problem of \textit{how} atomicity should be achieved. The literature on TM has been very prolific over the last decade, leading to the development of a plethora of solutions in software~\cite{Shavit1995STM,Felber10}, % 
hardware~\cite{Herlihy93, Yoo2013TSX, Nakaike2015IBM} 
and combinations thereof~\cite{Calciu2014a}. 
Existing TM solutions, however, consider homogeneous systems, in which threads execute either on CPUs~\cite{Yoo2013TSX,Riegel2007TinySTM} \textit{or} on (discrete) GPUs~\cite{Fung2012,Villegas15}. 
As such, existing TM systems fall short in harnessing the full potential of heterogeneous systems, failing to support execution scenarios in which CPUs and GPUs cooperate by concurrently accessing and manipulating the same state~\cite{Hetherington15,Zhong15,Mittal15} --- which is precisely the goal pursued by \abstraction.
Being an extension of the TM abstraction, \abstraction is particularly attractive for irregular applications, such as graphs and complex  data structures that make extensive use of pointers, for which designing scalable locking schemes can be complex. \abstraction enables the possibility of accelerating this class of applications by offloading part of the computations, which would be executed concurrently by some CPU threads, to the GPU.

%%PAO-CR: Kmeans and PS are not clear instances of the above mentioned classes of applications. I'll nuke it for the moment, may be can put it back in the conclusions as future work.
%Some concrete examples include K-means~\cite{Nelson19}
%(where CPU and GPU could process different data points and concurrently update the centroids) --- TODO: cannot add these insights in the intro
%or the Parameter Server model for machine learning~\cite{Li14}.
%(where the state of the model could be stored in STMR and be updated concurrently by worker threads running on both the CPU and GPU).
%}
Building an efficient TM for heterogeneous systems, though, is far from being a trivial task. In fact, in homogeneous platforms, where the TM abstraction is confined within the boundaries of a single processing device, e.g., a multi-core system or a discrete GPU,  conflict detection can be implemented via fast communication channels, e.g., the caches of a multi-core system. This is what allows  existing TM designs to incur limited overhead, even though they trigger conflict detection multiple times during transaction execution, possibly as frequently as upon each memory access of a transaction. The \abstraction abstraction, though, spans physically separated computational devices, which communicate
via channels, such as PCIe~\cite{pciexpress}, that are orders of magnitude slower than the ones assumed by conventional TM systems for homogeneous platforms. In these settings, thus, conventional TM approaches that impose multiple system-wide synchronizations along the critical path of execution of \textit{each} transaction would incur prohibitive overheads that would cripple performance.

In this work, we tackle this challenge by presenting \sys (Speculative \hetm), the first implementation of the proposed \hetm abstraction. \sys leverages a set of novel techniques that operate in synergy to effectively mask the latency of the inter-device communication bus.

\vspace{4pt}\noindent \textit{Hierarchical conflict detection.} \sys's novel hierarchical approach aims at removing inter-device conflict detection from the critical path of transaction's execution, in order to amortize its cost across batches of transactions.
More precisely, \sys detects \textbf{intra-device conflicts}, i.e., conflicts generated between transactions that execute on the same device (CPU or GPU), by relying on conventional TM implementations for homogeneous systems --- an approach that we term \textit{synchronous}, as conflicts are detected during transaction execution.
\textbf{Inter-device conflicts}, conversely, are checked \textit{asynchronously}, i.e., conflicts are detected periodically between \textit{batches} of transactions that are concurrently executed and committed, in a speculative fashion, at different devices. In absence of conflicts, the updates of each device are merged, yielding a consistent state at both devices. If inter-device conflicts are revealed, the speculatively committed transactions are rolled back and the state of the devices whose transactions were discarded is re-aligned to that of the ``winning'' device.
    
The use of speculation and asynchronous inter-device conflict detection not only amortizes the performance toll imposed by the synchronization over a high-latency inter-connection bus 
across a large number of transactions,
it also enables the use of  embarrassingly-parallel conflict detection schemes that, by operating on large transaction batches, can be very efficiently executed by modern GPUs.

\vspace{4pt}\noindent \textit{Non-blocking inter-device synchronization:} 
Although the cost of inter-device conflict detection can be amortized over a batch of transactions, the larger the batch of transactions processed in a synchronization round, the higher the likelihood of experiencing conflicts across devices. 
Thus, in conflict prone workloads, where using smaller transaction batches is desirable, it is crucial for reducing the overhead of the inter-device synchronization by minimizing the period of time during which transaction processing is blocked. 
To this end, \sys introduces an innovative scheme that ensures that, even while inter-device synchronization is being performed,  either the CPU or the GPU are  able to process transactions. This goal is achieved by combining two mechanisms: $i)$ overlapping the GPU-based validation of the transactions' batch with the processing of transactions on the CPU-side; $ii)$ letting the GPU start processing the transactions of the next synchronization round, while the  updates produced by the transactions it executed in the current round are being copied back to the CPU.

\vspace{4pt}\noindent \textit{Conflict-aware dispatching \& early validation:}  \sys exposes a programmatic interface that allows to control the assignment of transactions to either CPU or GPU. \sys exploits this information to implement a conflict-aware transaction dispatching scheme that aims at reducing the likelihood of inter-device contention.
%(which, as mentioned, leads to rolling-back a batch of speculatively committed transactions at either device)
This is achieved by dispatching transactions that are likely to contend
to the same device, where conflicts can be detected and resolved efficiently using the local TM implementation. 
Further, in case  inter-device contention does arise, \sys employs an early validation scheme that aims at reducing overheads (i.e., wasted work) by detecting conflicts before the synchronization for the current round is activated.  

\vspace{4pt}\noindent \textit{Modular and extensible design.} \sys is designed to ease integration with generic CPU-based and GPU-based TM implementations. To this end, \sys exposes a simple generic interface, which a TM needs to invoke in order to expose to \sys the read-sets and write-sets of the transactions it speculatively commits.
The ability of \sys to incorporate different TM implementations is quite relevant in practice, given that the design space of TM  is very wide and a number of studies have shown that no-one-size-fits-all TM implementation exists that can ensure optimal performance across all possible workloads~\cite{spear-adaptive,Didona2016ProteusTM}. This flexibility allows therefore to easily incorporate in \sys additional TM implementations, and to further increase the robustness of its performance in a wide spectrum of workloads.

\vspace{4pt} We evaluate \sys via an extensive experimental study, based  on synthetic benchmarks --- which we  use to shape workloads aimed at quantifying the overheads and gains derived from the various mechanisms \sys employs --- and a real world application,  MemcachedGPU~\cite{Hetherington15} --- which allows us to assess \sys's performance with a realistic workload as well as to showcase the benefits, in terms of load balancing and ease of programming, stemming from the possibility of concurrently accessing common data from physically separated computational units.

\section{Related Work}
\label{sec:rel_work}

Existing programming models for heterogeneous systems aim to provide different abstraction levels to unify the execution among devices with different architectures, programming paradigms and memory spaces. 
These models span from low-level and user-managed frameworks (such as OpenCL~\cite{openCL}) up to the fully automated run-time systems (e.g., StarPU~\cite{StarPU}, OmpSS~\cite{OmpSS} and Cashmere~\cite{Cashmere}). %% automatized
Other recent efforts aimed at simplifying the accelerator programming with high-level OpenMP-like directives, as highlighted in OpenACC~\cite{OpenACC} and OpenMP 4.0~\cite{OpenMP40}. %% evidenced
%Although some of these general frameworks predominantly focus on the functional code portability or may even include dedicated subsystems for implicit cross-device data management and task scheduling, 
%they do not necessary provide any guarantees regarding the performance portability and they may induce significant execution overheads. 
%For these reasons, the most widely used approach to fully exploit the performance of heterogeneous devices is still to combine the natively supported programming models on a per device type basis, such as CUDA along with OpenMP and Pthreads (an approach also adopted here for \abstraction). 
There are a number of ongoing efforts in the academia and industry aimed to automate data management and to unify memory in hybrid accelerated systems, for example, at the 
compiler level (CGCM~\cite{CGCM}, Spark-GPU~\cite{sparkGPU} and RSVM~\cite{RSVM}), NVIDIA CUDA Unified Memory~\cite{cudaGuide} or even support at the Linux kernel level~\cite{LinuxHMM}.
%Furthermore, one of the yet-to-be-solved burdens to fully unlock the potential of heterogeneous computing lies in the necessity of providing the automatic data management and support for unified memory in hybrid accelerated systems, 
%which is clear in many ongoing efforts from both academia and industry, %% evidenced
%such as compiler-level optimizations (ADSM, CGCM, Spark-GPU and RSVM), NVIDIA CUDA Unified Memory or even support at the OS kernel level, 
%e%.g., recently announced heterogeneous memory management (HMM) feature in Linux~\cite{LinuxHMM}. 
These solutions share our common high-level goal of simplifying the development of applications for heterogeneous systems. Yet, none of them tackle the challenges involved in ensuring the  consistency  guarantees provide by TM~\cite{Guerraoui08}, exposing programmers to the notorious complexity of lock-based synchronization~\cite{composableTM}. 

As mentioned, the literature in the area of TM has elaborated a plethora of design, exploring both hardware and software implementations. Although the majority of the existing literature focus on TM implementations for CPUs, TM is also gaining space in the GPU world~\cite{Fung2012,Holey2014,Xu14,Shen15}. In this area, a relevant related work is the recent  APUTM~\cite{Villegas2018}, which addressed the problem of implementing a STM for integrated GPUs. However, integrated GPUs reside in the same coherent domain as the CPU, unlike the case of discrete GPUs --- which we target in our work. As such, developing a TM for integrated GPUs is a much less challenging endeavor, as, in fact, this problem can be solved  by re-using existing designs for CPU TMs. To the best of our knowledge, our work is the first to present a TM system for heterogeneous systems that encompass both CPUs and discrete GPUs. It is also the first  work to revisit the definition of conventional TM  consistency semantics, e.g.,~\cite{Guerraoui08,vwc}, to  keep into account the specific architectural characteristics of  heterogeneous systems.

In a broad sense, \hetm is related to the work on speculative processing in distributed systems. 
In particular, optimistic simulation systems~\cite{Fujimoto-1990, Pellegrini-2011}, where the state of local simulation objects is allowed to advance in a speculative fashion, i.e.,~skipping synchronization with remote objects and rolling back to a consistent state if \emph{a posteriori} it is detected to have missed any relevant event from a remote object. 
Another related area has been investigated for speculative transaction processing techniques in distributed and replicated databases~\cite{specula,STR}.
Similarly, in this case the principle is similar, letting transactions commit speculatively and automatically roll-back the state of individual database replicas in case of any errors in speculation. 
\hetm builds on the same principles, but introduces new ad hoc designed techniques to meet the characteristics of heterogeneous systems composed of GPU and CPU.

\section{Defining the \hetm abstraction}
% \section{Design \& Model of HeTM}
% \section{System overview}
\label{sec:sys-arch}
\label{sec:abstraction}

As mentioned, \hetm provides the illusion of a single transactional shared memory that is concurrently accessed by a set of physically separated devices, where devices are equipped with their own local memory and  communicate over an interconnection bus like PCIe.  

In the definition of the \hetm abstraction we do not consider, for the sake of generality,  how transactions are generated and dispatched to the various execution devices. We leave the definition of these aspects to concrete implementation of the \hetm abstraction (see Section~\ref{sec:sys}). We will simply assume that threads, in execution at any computational device attached to the \hetm platform, can access and manipulate its state exclusively by means of transactions. To this end, \hetm exposes a conventional API, through which  threads can start a new transaction, submit read and write operations and request the commit or abort of the transaction. Extending the proposed \hetm abstraction to support intra-transaction parallelism~\cite{parallelNesting,futuresTM} and non-transactional accesses~\cite{strongatomicity} would be possible, but it  is outside of the scope of this work.

The rest of this section focuses on defining the correctness semantics that should be expected from a \hetm platform, 
such as the one that we will present in Section~\ref{sec:sys}, 
that exploits speculative techniques to mask the costs of inter-device synchronization. 
More in detail, we intend to reason on the correctness of TM implementations that can commit transactions in a speculative fashion, i.e., without first checking for conflicts with transactions executing on remote devices, and that may therefore have to be later aborted in case an inter-device contention is eventually detected. 

This speculative transaction execution model --- in which transactions are first \textit{speculatively committed} based only on local information, and only subsequently are \textit{committed} --- is desirable, in a \hetm platform, as it allows to remove intra-device conflict detection from the critical path of transactions' execution. In fact, in such a model, upon a transaction $T$ is speculatively committed, $i)$ the thread that requested $T$'s commit can be unblocked and process new transactions, and $ii)$ $T$'s updates can be immediately made visible locally.
%to other local transactions.
On the other hand, this  speculative execution model also enables  a broader spectrum of concurrency anomalies with respect to conventional transaction execution models that do not contemplate the notion of speculatively committed transactions.

We start by observing that existing consistency criteria  for classical TM systems, such as Opacity and Virtual World Consistency~\cite{Guerraoui08,vwc} (see Section~\ref{sec:rel_work}), are unfit to capture the dynamics of the speculative transaction execution model that we advocate to enable efficient implementations of the \hetm abstraction.  Roughly speaking, existing TM consistency criteria ensure, with various nuances,  two key properties:
\begin{itemize}

\item {P1. \textit{The behavior of every committed transaction has to be justifiable by the \textit{same} sequential execution containing only committed transactions,  without contradicting real-time order.}}

\item {P2. \textit{The behavior of any active transaction, even if it eventually aborts, has to be justifiable by \textit{some} sequential execution (possibly different) containing only committed transactions.}}

\end{itemize}
We argue that property P1
%which specifies the correctness semantics of committed transactions
remains adequate for a \hetm system. In fact, to preserve the ease of use of the TM abstraction, speculation should serve solely to enhance efficiency, while being totally hidden to applications. As such, the consistency semantics of committed transactions should remain unaltered, even if speculation is used for efficiency reasons.
 
Property P2, on the other hand, appears  unfit  to define the consistency semantics of \hetm platforms. In fact, the specification of P2 prohibits  observing the updates of \textit{any} uncommitted transaction, thus including the updates of speculatively committed transactions. Hence, if a transaction $T$ attempted to read a  data item updated by a speculatively committed transaction $T'$,  P2 would oblige any \hetm implementation to block $T$ until the final outcome (commit/abort) of $T'$ is determined --- limiting  the effectiveness of speculation to mask the costs of inter-device synchronization. 
 
Note also that allowing active transactions to observe the effects of \textit{any} speculatively committed transaction would not be a viable solution either. In fact, it would allow a transaction to observe the effects of two \textit{conflicting} speculatively committed transactions. This  would defeat the motivation at the basis of P2: avoiding  that applications may fail in complex/unpredicted ways due to  observing a state that no sequential execution could have ever produced.

Overall, we argue that consistency semantics offered by a \hetm platform should depart from classical consistency TM criteria by allowing different devices to use different sequential transaction histories to justify the execution of their local transactions. Intuitively,  these transaction histories should be composed by: $i)$ a prefix (possibly of different size) of the sequential execution history containing committed transactions (which, by P1, must be the same at each device), followed by $ii)$ a device-dependent sequential history composed by  transactions that speculatively-committed at that device.

We capture these semantics via the variant of property P2:
\begin{itemize}
\item {P2$^\dagger$. \textit{The behavior of any active or speculatively committed transaction $T$ has to be justifiable by  \textit{some} sequential execution containing $i)$ committed transactions and $ii)$ speculatively committed transactions that executed on the same device as $T$.}}
\end{itemize}

Properties P2$^\dagger$ and P2 pursue the same high-level goal: guaranteeing that the state observed by any transaction $T$ could have been produced in some sequential execution. Unlike P2, though, P2$^\dagger$  allows to include in the sequential execution used to justify $T$'s execution not only committed transactions, but also speculatively committed transactions that executed on the same device as $T$. This means that  transactions that execute at different devices must  observe a common history of committed transactions, but may witness the effects of  different speculatively committed transactions, which are still being checked for inter-device conflicts.

Note that P2$^\dagger$ also requires that the behavior of speculatively committed transactions (and not just that of active transactions) can be justified by a sequential execution. As active transactions can only read from committed or speculatively committed transactions, this implies that the only updates that can ever be observed are the ones produced by transactions that reflect some sequential history.
Further, a transaction $T$ can observe the effects of a (speculatively committed or committed) transaction $T'$, only provided that $T'$ does not conflict with any other transaction $T''$, whose effects $T$ has already observed so far. In fact, if $T$ were to observe the effects of two transactions that conflict either directly or indirectly, it would be impossible to include them both in the same sequential execution history that should be used to justify the execution of $T$.

\section{The \sys platform}
\label{sec:sys}

This section presents SHeTM (Speculative \hetm), an implementation of the \hetm abstraction that relies on speculation to minimize the overheads of inter-device synchronization. 

We start by describing the architecture and API that \sys exposes to application developers (Section~\ref{sec:sys:arch}), as well as the requirements for integration with guest TM libraries (Section~\ref{sec:sys:integration}). We also expose the basic functional principle behind the proposed solution in Section ~\ref{sec:sys:basic}, followed by a set of optimization techniques employed to boost its performance (Section~\ref{sec:sys:optim}. Finally, Section~\ref{sec:sys:policy} elaborates additional policies for conflict resolution. 

\subsection{Architecture and programming model}
\label{sec:sys:arch}

\sys implements the proposed \hetm abstraction for heterogeneous platforms composed by one or more cache-coherent multi-core CPUs and a discrete GPU. \sys is implemented in C and relies on the CUDA API to orchestrate the execution of the GPU. 

\sys maintains  a full replica of the shared TM region, which we call \mem (Speculative Transactional Memory Region), on both the CPU and GPU. At each device, the execution of transactions is regulated by a local TM library, referred to as \textit{guest} libraries. \sys adopts a  modular software architecture that seeks to attain inter-operability with generic TM implementations for CPU and GPU. This feature is important, since supporting the integration of arbitrary guest TM libraries allows to adapt the choice of the TM implementation used on each device to the characteristics of the application workload and the device. In Section~\ref{sec:arc:integration} we discuss which mechanisms \sys employs to integrate third-party guest TM libraries, as well as the assumptions that these libraries need to satisfy to correctly inter-operate with \sys.

\vspace{4pt} \noindent \textbf{Programming model.} \sys offers a conventional TM interface for demarcating  (i.e., beginning, committing, aborting) transactions and declaring read/write accesses to the \mem.  There are, however, relevant aspects related to the heterogeneous nature of the \hetm abstraction that programmers should take into account when developing transactional applications for \sys and that have influenced the design of \sys programming interfaces.

A first observation is that the \mem's replicas maintained by the CPU and GPU may be mapped in different positions in their address spaces. Thus, the management of pointers in \sys raises issues analogous to the ones that affect other implementations of the shared memory abstraction (e.g., in POSIX mmap or SystemV shmem~\cite{osbook}), such as: if pointers to a position within the \mem are stored in the \mem, they must be expressed as relative offsets and not as absolute addresses.
%(and converted back to absolute addresses before they are de-referenced).
%although \sys does not prevent storing in the \mem pointers to  memory regions external to the \mem, it is responsibility of the application developers to ensure that these pointers are only dereferenced by the device in whose address space they are defined.

A second relevant observation is that architectural differences of CPUs and (discrete) GPUs have a great impact  on their programming models and, as such,  \hetm systems should keep these aspects into account to attain high efficiency. One key issue is that, differently from CPUs, where transactions are typically executed individually, in GPUs it is desirable to execute transactions in relatively large batches~\cite{Shen15,cudaGuide}, as this allows for: $i)$  amortizing the latency of transactions' activation; $ii)$  enhancing throughput when transferring to/from the GPU the inputs/output required/produced by transactions' execution; $iii)$  improving resource utilization on modern GPUs.

To reconcile these differences, \sys abstracts over the computational model of CPU and GPU via a thread pool model in which each device exposes a number of worker threads. Worker threads are the only entities that can directly access the \mem, i.e., application threads that need to manipulate (or access) \mem should do so by submitting transactional requests to the worker threads (via \sys's API).
    
\sys views each instance of a transaction as an abstract operation that consumes an input and produces an output. \sys is opaque to the structure of transactions' inputs and outputs, requiring only information on their size in order to correctly transfer transactional requests/responses to/from the worker threads.
In order to support the efficient execution of transactions on both CPU and GPU, \sys allows programmers to associate each transaction via: $i)$ a ``transactional function'', which is meant to execute on the CPU and processes exactly one transaction; $ii)$ a ``transactional kernel'', which is meant to execute on the  GPU and processes a batch of transactions of a given size. 

Developers of transactional kernels have the responsibility to control which and how many threads to activate, how many transactions each thread should execute,  as well as how transactional inputs should be consumed. 
\sys, in turn, is responsible for activating transactional kernels, shipping to the GPU the corresponding transactions' inputs and retrieving the transactions' result to the host once the kernel ends. 

Programmers are not obliged to provide two implementations for a given transaction. 
If they do so, though, this provides \sys with the flexibility to select the implementation/device to use for executing a given transaction instance in a dynamic fashion, using a work-stealing policy that aims to balance load on both CPU and GPU. 

\vspace{4pt} \noindent \textbf{Transaction scheduling and dispatching.} For each registered transaction, \sys allocates a number of request queues. The number of queues that \sys allocates for a given transaction depends on the number of implementations that were registered for it. If a single implementation was defined (either for CPU or for GPU), only a single queue is created, which is used to store all the requests for that transaction. If implementations for both the CPU and GPU are provided, instead, \sys allocates three request queues, noted $\text{CPU}_Q$, $\text{GPU}_Q$ and $\text{SHARED}_Q$.  As their names suggest, the first two queues are meant to buffer requests which were submitted for execution on the CPU and GPU, respectively. This indication is passed to \sys via the programming interface used to support the submission of transactional requests, through which an optional \textit{device-affinity} parameter can be specified. 

This mechanism allows \sys to exploit external knowledge, e.g., provided  by programmers or  automatic tools (e.g., static code analysis~\cite{Riegel08} or on-line scheduling techniques~\cite{Diegues15, Shrink}), on the conflict patterns between different transaction instances and mitigate inter-device contention by dispatching conflict-prone transactions to the same device. %(where contention can be detected and managed more efficiently).

If, upon submission of a request, no \textit{device-affinity} is indicated (and both CPU and GPU implementations exist for the corresponding transaction), then the request is routed to $\text{SHARED}_Q$, which is accessible by both devices on the basis of a work-stealing policy.
Note that the enqueued requests are consumed at different granularity by the CPU and GPU. CPU worker threads process requests individually, extracting them from the $\text{CPU}_Q$ queues in a round-robin fashion, or, if $\text{CPU}_Q$ is found empty, from $\text{SHARED}_Q$. The processing of requests from the $\text{GPU}_Q$ queues, and the activation of the corresponding transactional kernel is coordinated by a management thread, which we call GPU-controller, running on the CPU. This thread monitors $\text{GPU}_Q$ and activates the corresponding transactional kernel when any $\text{GPU}_Q$ queue contains a sufficient number of requests to feed the kernel.
    
In many applications and standard benchmarks, transactions are naturally distributed to threads via some form of queue. This is the case of, e.g., MemcachedGPU~\cite{Hetherington15}, where transactions are triggered by the reception of network messages that are first stored into queues. Other examples are Intruder and Labyrinth of the STAMP benchmark suite~\cite{Minh08}. This type of applications naturally fits the programming model of \abstraction and incur no additional overheads due to the \sys's queuing system.

In applications that do not rely intrinsically on queues, the overhead of \sys varies depending on the workload characteristics: the larger the transaction execution time, the lower the overhead of the queuing system --- as the lower will be the frequency of access to the queue by the worker threads and, consequently, the likelihood of contention on the queue(s). Note, though, that, on the CPU side, programmers can bypass the \sys queuing system and let application level threads manipulate the \mem, provided that: $i)$ the \mem is only accessed transactionally via the same API used by the worker threads; and, $ii)$ the conflict resolution policy never aborts speculatively committed transactions on the CPU. In these conditions, correctness is preserved at the cost of exposing additional complexity to programmers --- as they become responsible for implementing the transaction scheduling and dispatching mechanisms provided by the \sys framework.
%\abstraction does not impose the usage of queues, CPU transactions could be executed directly on the STMR (the GPU would still require some batching scheme). This paper studies the situation where \sys has always transactions available, thus, collecting and enqueueing transactions introduces overheads that are out of its scope.

\subsection{Integration with guest TM libraries.} \label{sec:arc:integration}
\label{sec:sys:integration}

To guarantee the \hetm's consistency semantics described in Section~\ref{sec:abstraction}, \sys assumes that the underlying TMs ensure opacity 
(or, in general, any TM consistency criterion that guarantees the properties P1 and P2 defined in Section~\ref{sec:abstraction}).

\sys abstracts over the internal logic  of the guest TM libraries and interfaces with them by exposing a simple callback function that the guest TM should invoke, whenever a transaction commits, detecting conflicts in a hierarchical fashion, i.e., first locally in each device and then globally. 
%In a nutshell, the callback allows  \sys to detect inter-device transaction conflicts and keep track of the \mem updates produced at each device.
%The information communicated, via the callback function, by the guest TM to \sys differs for the CPU and GPU side.

\vspace{4pt} \noindent \textbf{CPU instrumentation.} On the CPU side, upon the commit of a transaction, a guest TM library must provide as input to \sys's callback function an array containing the $\langle \textit{address}, \textit{value}, \textit{timestamp} \rangle $ of each memory position \textit{updated} by that transaction. The specified timestamp must be usable by \hetm to totally order the updates to that memory position and is easily provided both by software and hardware TM implementations. For instance, most software TM implementations, e.g., TinySTM~\cite{Felber10} or NoREC~\cite{Dalessandro10},  use a logical timestamp to totally order the commits of all transactions.
The same can be done in hardware TM implementations, such as Intel TSX, by reading the processor cycles, i.e., using the RDTSCP instruction~\cite{Castro18}.
%With hardware TM implementations, such as Intel TSX, the processor cycles (e.g., read via the RDTSCP instruction) can be used to determine the total order. 
Gathering  transactions' write-sets  imposes no additional overhead to a guest STM, as STMs need anyway to track the write-sets in software. For HTM, \sys requires the software instrumentation of write operations to gather the transaction's write-set. It is worth noting that, in many realistic workloads, writes are largely outnumbered by reads and, as such, the resulting instrumentation overhead is small.
The \hetm's callback function appends the write-sets into thread-local data-structures, referred herein as the \textbf{CPU write-set logs}, and periodically offloads them to the GPU to perform inter-device conflict detection.

\vspace{4pt} \noindent \textbf{GPU instrumentation.} On the GPU side, a guest TM library must communicate to the \hetm's callback the set of addresses read and written by the committing transaction. 
%In the CPU case,  the write-sets are accumulated in the per thread logs until inter-device synchronization is activated. 
Conversely, on the GPU side the read-set and write-set of a committing transactions are used to update two bitmaps, noted \textbf{$\text{RS}^\text{GPU}_\text{bmp}$} and \textbf{$\text{WS}^\text{GPU}_\text{bmp}$}, that track the regions of the \mem that GPU transactions read or wrote, respectively.
After those bitmaps are updated, the transaction's read-set and write-set can be immediately discarded. The necessity for this asymmetric instrumentation logic at the CPU and GPU is further detailed in Section~\ref{sec:sys:basic}.

\vspace{4pt} \noindent \textbf{Additional assumptions.}
\sys needs to manipulate the state of the \mem to merge the updates produced at both devices and to cancel the effects of speculatively committed transactions in case  inter-device conflicts are detected. These updates are performed in a non-transactional way, i.e., bypassing the APIs of the guest TM library --- ensuring that no transaction is executing concurrently, to preserve consistency. This design is safe under the assumption that any meta-data managed by the guest TM libraries are maintained externally to the \mem, e.g., in a disjoint memory region on the local device. 
This assumption is met in practice by most TM implementations, and is valid for all existing HTM implementations (which maintain their metadata in the processor's caches),
and for all word-based STMs (where TM metadata must necessarily be stored in a disjoint memory region to avoid interference with the application's memory layout).

\vspace{4pt} \noindent \textbf{Supported libraries.} 
Currently, \sys  supports three TM implementations: two on the CPU side -- TinySTM~\cite{Felber10} and Intel's TSX~\cite{IntelCorporation2015}, implemented respectively in software and hardware -- and one on the GPU side, namely PR-STM~\cite{Shen15}.
    
\begin{figure*}[tbp]
	\centering
	\begin{subfigure}[b]{0.49\textwidth}
	\begin{subfigure}[b]{0.4\textwidth}
	\caption*{Case of commit.}
	\includegraphics[height=2.2cm]{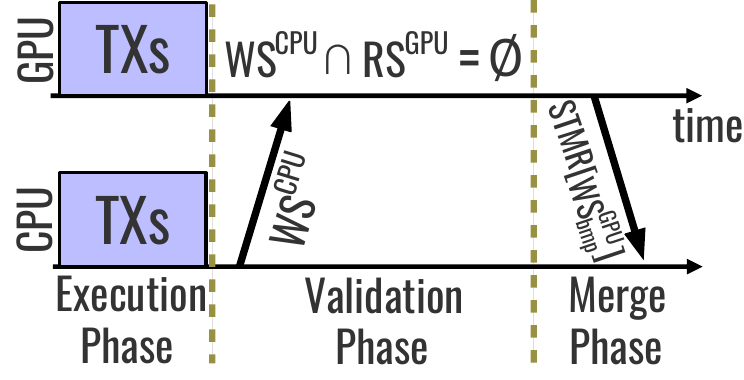}
	\vspace{-0.3cm}
	\end{subfigure}
	\quad\quad
	\begin{subfigure}[b]{0.4\textwidth}
	\caption*{Case of abort.}
	\includegraphics[height=2.2cm]{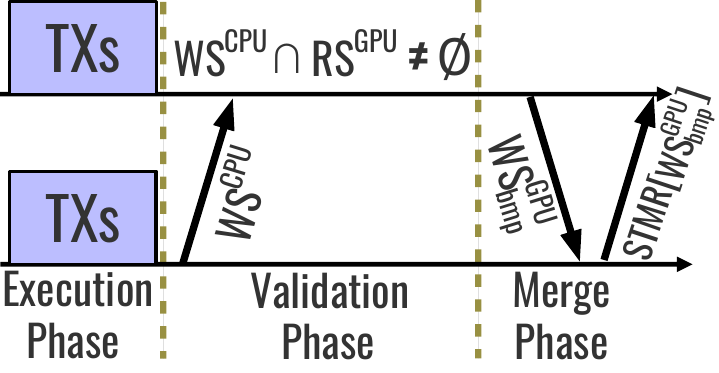}
	\vspace{-0.3cm}
	\end{subfigure}
	\vspace{-0.1cm}
	\caption{A basic variant of \sys.}
	\label{fig:sys_hetm_simple}
	\end{subfigure}
	\hfill
	\begin{subfigure}[b]{0.49\textwidth}
	\begin{subfigure}[b]{0.4\textwidth}
	\caption*{Case of commit.}
	\includegraphics[height=2.7cm]{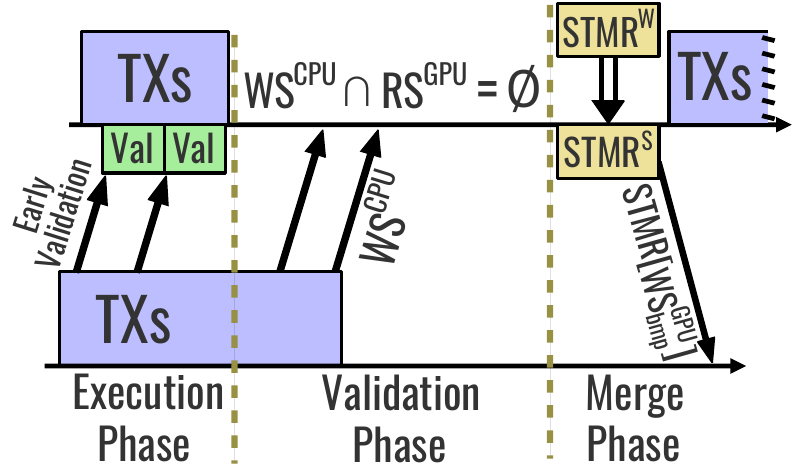}
	\vspace{-0.3cm}
	\end{subfigure}
	\quad\quad
	\begin{subfigure}[b]{0.4\textwidth}
	\caption*{Case of abort.}
	\includegraphics[height=2.7cm]{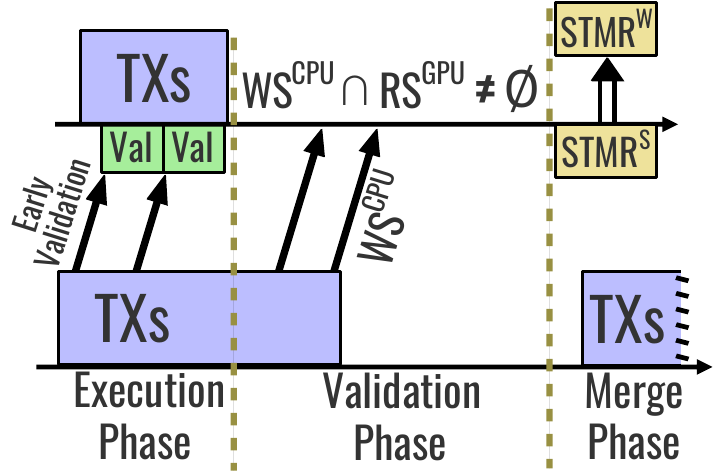}
	\vspace{-0.3cm}
	\end{subfigure}
	\vspace{-0.1cm}
	\caption{Illustrating the behavior of \sys.}
	\label{fig:sys_hetm}
	\end{subfigure}
	\vspace{-0.1cm}
	\caption{Figure~\ref{fig:sys_hetm_simple} illustrates a simple variant of \sys, while Figure~\ref{fig:sys_hetm} presents how to avoid unnecessary copies.}
	\vspace{-0.5cm}
\end{figure*}

\subsection{Basic Algorithm}
\label{sec:sys:basic}

We start by describing a basic variant of the \sys's algorithm that serves a twofold purpose: $i)$ it allows us to simplify presentation, by explaining the design of  \sys  in an incremental fashion; and, $ii)$ it exposes several sources of inefficiency that we address in the following text.
At this stage, we will  assume a fixed policy to deal with inter-device contention that deterministically discards the transactions speculatively committed by the GPU.
A discussion on how to relax this assumption and support policies that discard transactions speculatively committed by the CPU can be found in the Section~\ref{sec:sys:policy}.

\sys orchestrates the execution of GPU and CPU in \textit{synchronization rounds}, where each  round is composed by three phases: execution, validation, and merge (see Figure~\ref{fig:sys_hetm_simple}). In a nutshell, \sys adopts a \textit{hierarchical conflict detection mechanism} that operates as follows: during the execution phase, conflicts are detected  only among transactions that execute on the same device, using a local TM implementation; inter-device conflicts are then verified during the validation phase via a novel scheme that leverages the massive parallelism of modern GPUs to maximize performance.

\subsubsection{\textbf{Execution Phase}}

In the execution phase, transactions are extracted from the input queues and fed to the devices during a user-tunable period. Transaction processing is executed in an independent way at both devices, starting from a consistent snapshot, i.e., an identical replica of the \mem at both devices, and executing transactions in a speculative fashion: the execution of transactions is regulated exclusively by the local TM library, which only detects conflicts between local transactions, avoiding any inter-device synchronization.

When a transaction requests to commit, the unmodified commit logic of the local TM library is used to atomically propagate the transaction's update to the local \mem replica. This local commit event coincides with the speculative commit, in the execution model assumed by the \hetm abstraction. At this point the TM library invokes the call-back functions exposed by \sys, as referred in Section~\ref{sec:sys:integration}. On the CPU, the write-set of the transaction is appended to a per-thread log. On the GPU side, the transaction's read-set and write-set are used to update $\text{RS}^\text{GPU}_\text{bmp}$ and $\text{WS}^\text{GPU}_\text{bmp}$ bitmaps. The  bitmaps encode the set of addresses read and written by every  transaction that speculatively committed during the execution phase, and they are updated concurrently by the GPU threads that are in charge of executing transactions.  
        
As mentioned, the duration of execution phase is a user-tunable parameter that allows to explore an interesting performance trade-off, which will be studied in Section~\ref{sec:eval:memcached}. Longer periods imply less frequent synchronizations, which means lower overhead in case the synchronization is successful. However, longer period of executions mean also that a larger number of transactions are speculatively executed at both devices, increasing the probability of inter-device contention --- thus leading to wasting more work (aborted transactions). 

\subsubsection{\textbf{Validation Phase}}

The goal of this stage is to determine whether there was any conflict between the transactions processed by the CPU and the GPU during the execution phase. We designed the logic of this phase on the basis of the following key observations:

\vspace{4pt} \noindent $\bullet$ As the local TM libraries are assumed to ensure opacity, the behavior of the speculatively committed transactions at each device is already guaranteed to be equivalent to a sequential execution (although defined over different sets of transactions). The set $S$ of transactions speculatively committed during the processing phase at a device $D\in\{\text{CPU},\text{GPU}\}$ can thus be logically subsumed by a single, equivalent transaction, noted $T^D$, whose read-set and write-set is the union of the read-sets and write-sets of the transactions in $S$.  This observation allows us to reduce  the problem of conflict detection among a number of speculatively committed transactions to the problem of detecting conflicts between a pair of logical equivalent transactions, which we denote as $T^\text{CPU}$ and $T^\text{GPU}$.

%that, if no conflicts are detected between the  equivalent transactions $T^\text{CPU}$ and $T^\text{GPU}$, their effects can be merged by applying the write-set of the logical transaction executed on the CPU on the GPU-side, and vice-versa.
            
\vspace{4pt} \noindent $\bullet$ Detecting conflicts between a pair of  transactions  can be reduced to verifying intersections between their read-sets and write-sets~\cite{Papadimitriou79, Bernstein1986}. This computation can be efficiently parallelized using GPUs, especially if the sets are large, as it is the case for $T^\text{CPU}$ and $T^\text{GPU}$, which subsume a (typically very large) number of transactions.

The design of the \sys's validation scheme, as well  as its instrumentation logic,   were engineered, based on this observation, so as to take full advantage of modern GPUs.
             
\vspace{4pt} \noindent $\bullet$  In many realistic workloads, transactions read a much larger number of memory positions than they write to~\cite{Felber2016}. % ShadyEurosys16-and-more-can-be-found-in-that-paper 
As such, the read-sets of $T^\text{CPU}$ and $T^\text{GPU}$ are likely to be much larger than their corresponding write-sets. Motivated by this observation, we designed the validation scheme in a way to avoid transmitting the read-sets over the inter-connection bus.
Note that there are two possible orders in which  $T^\text{CPU}$ and $T^\text{GPU}$ may be serialized, namely  $T^\text{GPU}\rightarrow T^\text{CPU}$ or $T^\text{CPU}\rightarrow T^\text{GPU}$. For the former order to be valid, none of the writes generated by $T^\text{GPU}$ should be ``missed'' by $T^\text{CPU}$, i.e., $\text{WS}^\text{GPU}\cap \text{RS}^\text{CPU} = \emptyset$ (where $\text{WS}^\text{GPU}$ and $\text{RS}^\text{CPU}$ denote the write-set and read-set of $T^\text{GPU}$ and $T^\text{CPU}$, respectively). The latter order, conversely, requires verifying whether $\text{WS}^\text{CPU}\cap \text{RS}^\text{GPU}=\emptyset$.
Keeping into account that we intend to leverage the GPU to perform validation, 
\sys opts for testing whether the \textbf{CPU transactions can be serialized before the GPU transactions, i.e., $T^\text{CPU}\rightarrow T^\text{GPU}$}. 
In fact, the computation of  $\text{WS}^\text{CPU} \cap \text{RS}^\text{GPU} = \emptyset$ can be performed on the GPU side by shipping only the write-sets of the CPU transactions --- 
whereas, the opposite serialization order ($T^\text{GPU} \rightarrow T^\text{CPU}$) would require shipping to the GPU the read-sets of the CPU transactions.

\vspace{4pt} \noindent $\bullet$ In order to guarantee that the updates of the $T^\text{CPU}$ and $T^\text{GPU}$ can be mutually exchanged and applied at each device, yielding a state equivalent to the one produced by the schedule $T^\text{CPU} \rightarrow T^\text{GPU}$, 
it is necessary to exclude also the presence of write-write conflicts, i.e.,  whether  $\text{WS}^\text{CPU} \cap \text{WS}^\text{GPU}=\emptyset$.
The approach taken in \sys to guarantee this property is to track the writes performed by GPU transactions not only in $\text{WS}^\text{GPU}_\text{{bmp}}$, but also in $\text{RS}^\text{GPU}_\text{{bmp}}$. By guaranteeing that $\text{WS}^\text{GPU}\subseteq \text{RS}^\text{GPU}$, the verification of $\text{WS}^\text{CPU} \cap \text{RS}^\text{GPU}=\emptyset$ also ensures that   $\text{WS}^\text{CPU} \cap  \text{WS}^\text{GPU}=\emptyset$.
%This check, though, is necessary only if one assumes that transactions can issue ``blind writes'', i.e., writing to a memory position without reading it. 
%If blind writes can be excluded, in fact, the write-set of $T^\text{GPU}$ is guaranteed to be a subset of $\text{RS}^\text{GPU}$  ($\text{WS}^\text{GPU} \subseteq \text{RS}^\text{GPU}$). 
%As such, verifying that $\text{WS}^\text{GPU} \cap \text{RS}^\text{CPU}=\emptyset$ implies that  $\text{WS}^\text{GPU} \cap  \text{WS}^\text{CPU}=\emptyset$. 
%In transactional systems blind writes are considered as quite rare~\cite{Bernstein1986,Ullman1983Databses}. % someotherDBMStextbook
%Nonetheless, in \sys, we take a safe approach, whose correctness does not hinge on the absence of blind writes, but  still allows for sparing the cost of detecting write-write conflicts: 
%we guarantee that $\text{WS}^\text{GPU}\subseteq \text{RS}^\text{GPU}$ by tracking the writes issued by GPU transactions, not only in the write-set bitmap, but also in the read-set bitmap.
%
Furthermore, it is worth noting that writes are typically outnumbered by reads, the overhead incurred by tracking the writes in two bitmaps is expected to be low.

\vspace{4pt} Let us  put all these pieces together and discuss how the validation phase operates in a systematic fashion. 

The validation phase starts by transferring  the write-set logs gathered  by each CPU thread to the GPU. The logs are streamed in chunks to achieve high throughput  and activate validation kernels that operate at sufficient granularity to achieve high utilization of GPU resources. A validation kernel on the GPU takes as input a chunk of a log and operates as follows: For each tuple $ \langle \textit{address}, \textit{value}, \textit{timestamp} \rangle $ in the input log it is checked whether the corresponding entry in the GPU's read-set bitmap  is set --- which indicates that some of the transactions speculatively committed by the GPU during the execution phase read that address. 
        
If the CPU write is found to have invalidated the read-set of $T^\text{GPU}$ the validation phase returns a  negative outcome, but it continues applying the full set of write-set logs sent by the CPU. This ensures that, at the end of the validation phase, the GPU's \mem incorporates all the effects of $T^\text{CPU}$. Thus, if in the merge phase, the state of the GPU's \mem needs to be re-aligned to the current state of the CPU's \mem, it is sufficient to simply undo the effects of the $T^\text{GPU}$. 
        
If the CPU write does not invalidate the read-set of $T^\text{GPU}$, the corresponding value stored in the CPU write-set log is applied to the GPU's \mem. 
This is performed non-transactionally, since the GPU is not processing transactions during the validation phase. 
However, since the CPU logs are validated in arbitrary order on the GPU side, 
before applying it is necessary to verify if the version currently present in the GPU's \mem is not more recent than one that is being applied. 
To this end, on the GPU, \sys maintains a timestamp array, denoted as $TS$, 
which has an entry per word of the \mem reserved to store the timestamps of the CPU writes applied during the validation phase. 
During validation, GPU threads consult the $TS$ to determine whether the write being validated reflects a more recent state than the one already present in the \mem, applying that log tuple only if this is the case.  Note that since concurrent GPU threads may be validating writes targeting the same address, the atomicity of the test for freshness and the value application is ensured via a lock implemented using the first bit of the corresponding $TS$ entry.

\subsubsection{\textbf{Merge Phase}}

The merge phase ensures that  the replicas of the \mem at the CPU and at the GPU are consistent, before starting the execution phase of the next synchronization round.
The way in which the states of the CPU and GPU  are realigned depends on the outcome of the validation phase.

If the validation phase is successful, i.e., no inter-device conflicts are detected, the GPU's replica of the \mem already incorporates the updates of the transactions that speculatively committed at both devices,
since recall that, during the validation phase, the GPU also applies the CPU write-sets into its local \mem replica. 
To this end, the GPU-controller thread fetches the GPU's write-set bitmap, which identifies the memory regions updated by the transactions speculatively committed by the GPU and activates the memory transfers to update in-place the \mem's replica on the CPU.

If the validation phase fails, the state of the CPU's \mem overrides the GPU's \mem. 
To this end, the GPU controller thread obtains the GPU's write-set bitmap and transfers the CPU's state over the memory regions marked as updated on the GPU's write-set bitmap, thus undoing any side-effect of the execution of transactions on the GPU side.

\subsection{Optimizations}
\label{sec:sys:optim}

\sys integrates a number of additional mechanisms that aim at tackling two main sources of inefficiency: 
\textbf{the blocking time} (i.e., the period during which transaction processing is blocked) due to inter-device synchronization, 
and the overhead imposed in case of inter-device contention. We present these techniques in the following text and illustrate them in Figure~\ref{fig:sys_hetm}. %TODO: ~\ref{fig:hetm_system-check}.

\vspace{4pt} \noindent \textbf{Inter-device synchronization.} As illustrated in Figure~\ref{fig:sys_hetm_simple}, in the basic algorithm presented in Section~\ref{sec:sys:basic}, transaction processing is blocked throughout the validation and merge phases both at the CPU and at the GPU. This is clearly undesirable for efficiency reasons, especially if one considers that, to reduce the likelihood of inter-device contention, it is desirable to use relatively short execution phases.

\sys tackles this issue by integrating mechanisms aimed at reducing the blocking time both at the CPU and at the GPU.

On the CPU side, during the validation phase, \sys allows the worker threads to continue processing transactions concurrently with the streaming of the logs accumulated during the execution phase. The CPU blocking time on the execution phase only occurs when a very few log chunks are left to be offloaded to the GPU. In practical settings, the speed at which logs can be transferred is higher than that at which new logs can be produced by the worker threads. Thus,  this mechanism   effectively overlaps transaction processing at the CPU side with the log transfers to the GPU, while generating a relatively little amount of additional logs to validate for the GPU.

On the GPU side, at the end of the merge phase, the basic algorithm blocks transaction processing  while transferring to the CPU  the memory regions updated by the GPU. This is done to ensure that the state of the GPU's \mem is not corrupted due to the execution of transactions while the device to host transfer is ongoing. \sys tackles this problem by employing a double buffering approach.
At the start of the execution phase, a \textit{shadow copy}  (\memShadow in Figure~\ref{fig:sys_hetm}) of the GPU's \mem (\memWork in Figure~\ref{fig:sys_hetm}) is created, via a device to device copy. As soon as \memShadow is created, GPU transaction processing can immediately resume, 
since \memShadow is isolated from the updates of transactions (which operate exclusively on the \memWork) and can be used to feed the device to host transfer.

\vspace{4pt} \noindent \textbf{Inter-device contention.} As discussed in Section~\ref{sec:sys:arch}, \sys's API allows exploiting external knowledge on transactions' conflict patterns (via the device affinity specified at transactions' submission time) to control the dispatching of transactions and reduce inter-device contention.
Besides  striving to reduce the likelihood of inter-device contention, \sys incorporates two additional mechanisms that aim at reducing two sources of overhead when inter-device conflicts do occur:

\vspace{2pt}\noindent $\bullet$ \textbf{Wasted work.} In the basic algorithm, conflicts are detected only at the end of the execution phase. 
This leads to wasting a large amount of work at the GPU, if a conflict is detected in the validation phase. We tackle this problem by introducing an early validation scheme that periodically transfers the CPU's  logs to the GPU, where they are validated (but not applied) while transactions are concurrently processed on both devices. As early validations are concurrent with transaction processing, it is still necessary to  validate all the write-set logs produced during the execution phase in the validation phase. Yet, by anticipating the detection of inter-device conflict, as we will see in Section~\ref{sec:eval}, early-validation can provide significant gains in contention prone workloads by reducing the time the GPU spends performing computations that are eventually discarded.
    
\vspace{2pt}\noindent $\bullet$  \textbf{Rollback latency.} Realigning the GPU's state to that of the CPU, in case of inter-device contention, imposes significant overhead in the basic algorithm. Every memory region updated by the GPU has to be copied from the CPU and, during this transfer, transaction processing is blocked at both devices. Fortunately, the availability of the shadow copy is of great help in this case. Recall that the shadow copy reflects a consistent state of the \mem, as at the beginning of the current synchronization round. Thus, in order to align the shadow copy to the current state of the CPU, it suffices to apply to it the CPU's write-set logs. 

\vspace{2pt}\noindent $\bullet$  \textbf{Enhancing memory transfer's throughput.}
In order to exploit PCIe bandwidth while transferring information,  transfers are performed in \textit{chunks} of relatively coarse granularity. To this end, the CPU write-set logs are shipped to the GPU using a granularity of 48 KB; also, the write-set bitmap on the GPU tracks updates  with a granularity of 16KB.

As a further optimization, the GPU-controller coalesces transfers of contiguous  chunks from the GPU to the CPU during the merge phase (in the case of no inter-device contention), as well as when performing the device to device copy from the shadow to the working copy of the \mem (in the case of inter-device contention).

%Memory transfer granularity is tuned to effectively exploit the PCIe bandwidth, while transferring: $1)$ CPU write-sets' logs to GPU (validation phase); and, $2)$ modified \mem regions from GPU to CPU (merge phase). We experimentally found that 16KB is a value that to provide robust performance for both.
%To further optimize transfers, the GPU-controller coalesces contiguous updated \mem chunks.}

%%%% TODO: commented for space

\subsection{Additional Conflict Resolution Policies}
\label{sec:sys:policy}

The solution presented so far, as well as the current prototype of \sys, assumes a fixed conflict resolution that deterministically aborts the speculatively committed transactions on the GPU side in case of inter-device conflict. Note that this property has the advantage of ensuring that the effects of speculatively committed transactions on the CPU side can be immediately considered as committed. Thus their results can be externalized to application-level results without incurring the latency of inter-device synchronization. This is a desirable property in practice, as arguably the CPU is the preferred device to execute latency sensitive transactions (considering that the processing of GPU transactions is burdened by the latency of kernel activation and result transfer).

Nonetheless, supporting alternative conflict resolution policies, i.e., that abort the transactions speculatively committed by the CPU, may be useful, e.g., to avoid starving the GPU or to favor the device that committed more transactions.

Extending the scheme described so far to favor the GPU and discard the effects of the CPU transaction is, from an algorithmic perspective, quite straightforward. During the validation phase, applying the CPU write-sets is only done conditionally to the successful outcome of the entire validation procedure. If a conflict is detected, the apply phase is skipped and a negative outcome is returned to the GPU-controller thread. Using a technique analogous to the one presented in Section~\ref{sec:sys:optim}, a shadow copy of the CPU's \mem (created at the beginning of the current execution phase) can be used to undo the effects of the speculatively committed  transactions (on the CPU side). Finally, the  memory regions updated by the  GPU can be applied (in chunks) to the CPU's \mem.

On the CPU side, a natural and lightweight way to create a shadow copy of the \mem consists in forking  the process that hosts the worker threads. This approach, widely used to implement efficient checkpointing scheme~\cite{Liu2017DudeTM,Castro18,Kamino-TX}, allows to exploit the Copy-On-Write mechanism of the OS, avoiding synchronous memory copies.

%PAO-CR: NUKED since we don't have it and we don't have much space 
A contention management mechanism to avoid GPU starvation can be implemented as follows: if the GPU is aborted more than a predetermined number of times, it only allows read-only transactions to execute on the CPU-side in the subsequent execution phase, postponing the execution of enqueued update transactions until the next round. It is easy to observe that the absence of update transactions at the CPU guarantees the successful validation in the next round.

\section{Evaluation}
\label{sec:eval}

\begin{figure}[tbp]
	\centering
	\includegraphics[width=.8\columnwidth]{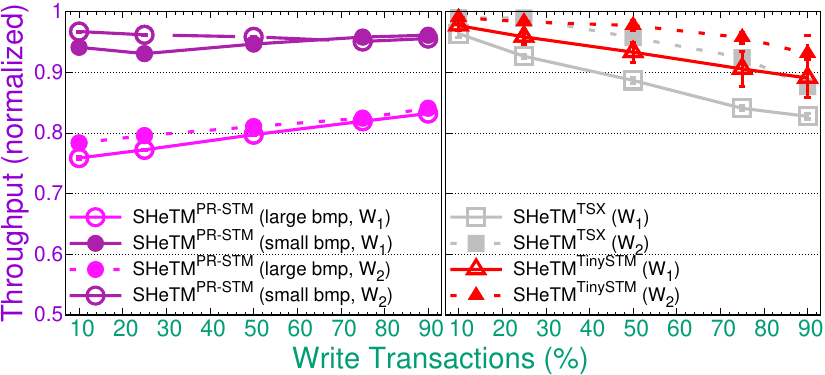}
	\vspace{-0.1cm}
	\caption{Cost of instrumentation of guest TM libraries.}
	\vspace{-0.5cm}
	\label{fig:instrCost}
\end{figure}

This Section presents an experimental study that aims at evaluating the following key aspects:
%answering the following key questions: 
the costs imposed by the instrumentation of the guest TM libraries (Sec.~\ref{sec:eval:instr}); overhead introduced to workloads whose scalability is not limited by inter-device contention (Sec.~\ref{sec:eval:scala}); performance degradation due to inter-device contention (Sec.~\ref{sec:eval:cont}); optimization gains over simpler designs (Sec.~\ref{sec:eval:scala} and Sec.~\ref{sec:eval:cont}); and finally, how effective \sys is with realistic applications (Sec.~\ref{sec:eval:memcached}). 

%This section presents an experimental study that aims at answering the following key questions:
%\begin{itemize}
    %\item What are the costs imposed by the instrumentation of guests TM libraries? (Sec.~\ref{sec:eval:instr})
    %\item What overheads does \sys introduce in workloads whose scalability is not limited by inter-device contention? (Sec.~\ref{sec:eval:scala})
    %\item How sensitive is the performance of \sys to inter-device contention? (Sec.~\ref{sec:eval:cont})
    %\item How large are the gains that \sys's optimizations enable over simpler designs? (Sec.~\ref{sec:eval:scala} and Sec.~\ref{sec:eval:cont})
    %\item How effective is \sys with realistic applications? (Sec.~\ref{sec:eval:memcached})
%\end{itemize}

%\subsection{System Specifications}

Our evaluation is conducted using a machine equipped with
an Intel Xeon E5-2648L v4 CPU (14 cores with support for HTM, 32GB DRAM), an Nvidia GTX 1080 GPU (8GB XDDR5, driver 387.34, CUDA 9.1), and running Ubuntu 16.04.3 LTS (kernel 4.4.0-57). Applications are manually instrumented to use the \sys API.

%The Nvidia driver's version is 387.34 and the CUDA framework's version is 9.1.

We based our evaluation on a set of synthetic benchmarks conceived to assess different aspects of \sys's design, and on MemcahedGPU~\cite{Hetherington15}.

In all the tests, we use 8 worker threads on the CPU side. As for the transactional kernels, we tuned their configuration (number of transactions per kernel activation, active threads and thread blocks) on the basis of preliminary evaluations to maximize the GPU throughput. The synthetic workloads use the same transactional logic on both the CPU and GPU and operate on a \mem of size 600MB, unless otherwise specified;
the \mem size in MemcachedGPU is around 480MB.

\subsection{Instrumentation Costs}
\label{sec:eval:instr}

Let us start by assessing the overhead induced by the software instrumentation that \sys requires for its guest TM libraries. To this end we consider two workloads, noted $W_1$ and $W_2$, that access the \mem uniformly at random. In $W_1$, read-only transactions issue 4 reads, whereas update transactions read and update 4 memory positions. $W_2$ is identical to $W_1$, except that both transaction types issue 40, and not 4 reads. $W_1$ is designed to stress the instrumentation of read and write operations. $W_2$ is selected as representative of many realistic workloads, in which reads outnumber the writes.

In the plot in Figure~\ref{fig:instrCost} we vary on the x-axis the percentage of the update transactions from 10\% to 90\% and report on the y-axis the throughput normalized w.r.t. un-instrumented version of PR-STM~\cite{Shen15} for the GPU (left plot), and of TinySTM~\cite{Felber10} and TSX for the CPU (right plot).

In the left plot (GPU), we consider using two different levels of tracking granularity for the read-set bitmap ($\text{RS}_\text{bmp}^\text{GPU}$), namely 4B (small bmp) and 1KB (large bmp). It can be observed that, independently of the considered workload, the use of the small granularity bitmaps induce, larger overheads, approx.~20\%, as its larger size leads to a lower locality of reference.
In contrast, the coarser granularity bitmap reduces significantly the instrumentation overhead, to approx.~5\%, at the cost, though, of spurious aborts due to the risk of false positives in the conflict detection scheme. In fact, the trade-off between instrumentation overhead and access tracking granularity is well known in the literature, e.g.,~TM~\cite{Felber10}.

In the right plot (CPU), we observe that the instrumentation cost is on average around 5\% for $W_2$ for both TinySTM and TSX. In all scenarios, the overhead is below 10\% except for  the most write intensive variants of  $W_1$, where it remains below 20\% even in presence of 90\% of update transactions.

\subsection{Efficiency in absence of inter-device contention}
\label{sec:eval:scala}

\begin{figure}[tbp]
	\centering
	\begin{subfigure}[b]{0.23\textwidth}
	\caption{100\% update transactions}
	\vspace{-0.25cm}
	\includegraphics[width=0.99\columnwidth]{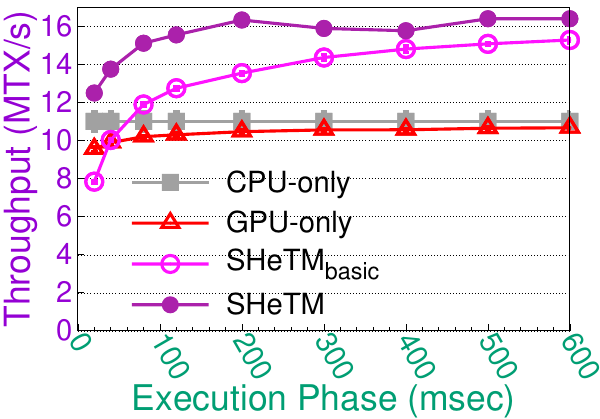}
	\end{subfigure}
	~
	\begin{subfigure}[b]{0.23\textwidth}
	\caption{10\% update transactions}
	\vspace{-0.25cm}
	\includegraphics[width=0.99\columnwidth]{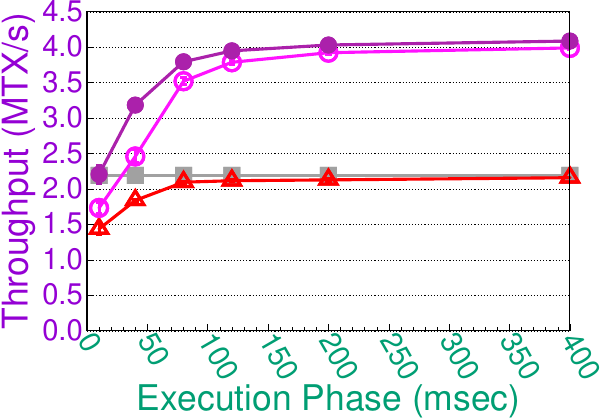}
	\end{subfigure}
	\vspace{-0.1cm}
	\caption{Efficiency in absence on contention.}
	\vspace{-0.5cm}
	\label{fig:overhead_cmp}
\end{figure}

Next, we intend to assess which overheads \sys incurs in workloads whose scalability is not limited by inter-device contention. Here, we consider two variants of the $W_1$ workload, generating 100\% ($W_1$-100\%) and 10\% ($W_1$-10\%) update transactions, respectively.

We avoid inter-device contention by partitioning the \mem in two halves and restricting CPU and GPU to access a different half.  The results of this study are reported in Figure~\ref{fig:overhead_cmp}, in which we vary on the x-axis the duration of the execution phase from 1 msec to 600 msec and report on the y-axis the throughput of \sys and of the following baselines: the basic variant of \sys presented in Section~\ref{sec:sys:basic}, noted\textit{ \sys basic}; TSX running solo, noted \textit{CPU-only}; PR-STM running solo and copying its \mem to the host, after executing a kernel, using double buffer (i.e., without blocking), noted \textit{GPU-only}.

\begin{figure}[tbp]
	\centering
	\includegraphics[width=1.0\columnwidth]{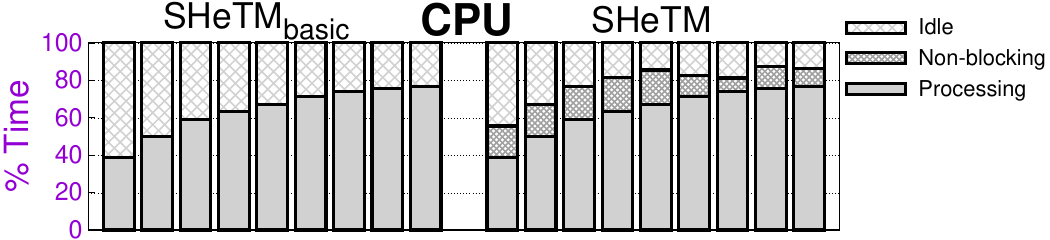}
	\includegraphics[width=1.0\columnwidth]{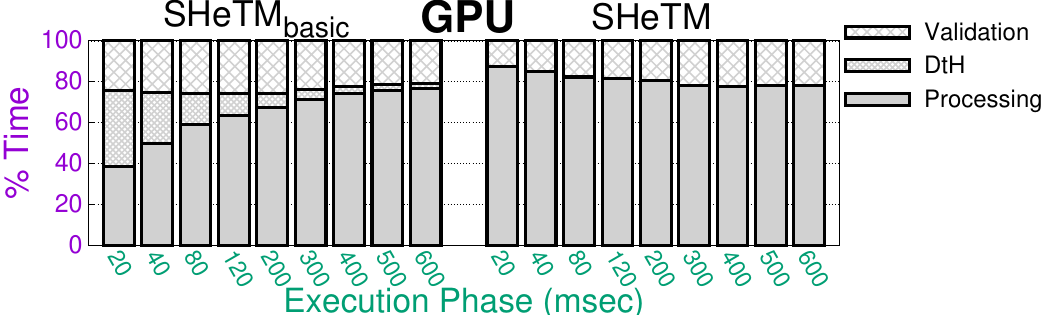}
	\vspace{-0.5cm}
	\caption{Break-down of exec. times (100\% update transactions)}
	\label{fig:overhead_time}
	\vspace{-0.7cm}
\end{figure}

The throughput plot on the left, which refers to $W_1$-100\%, shows that as the execution period grows the performance of \sys also increases, 
which is as expected, since the relative amount of time spent performing the validation and merge phases reduces, amortizing their cost over larger period of useful processing (see Figure~\ref{fig:overhead_time}). The peak throughput of approx. 17M tx/sec, is reached at 200 msecs and plateaus beyond that value. \sys's peak throughput is about 55\% higher than the peak throughput of CPU-only and GPU-only (approx.~11 M tx/sec) and only 23\% lower than the throughput of an idealized system that could total the combined throughput of both uninstrumented devices. 
 
 By contrasting the performance of \sys with that of \textit{basic} we can clearly appreciate the performance gains enabled by the optimizations described in Section~\ref{sec:sys:optim}, which are particularly significant with small execution periods  (up to +56\% higher throughput when the execution period lasts 1 msec). The bar plots in Figure~\ref{fig:overhead_time}, which report the breakdown of times spent by the CPU and GPU in various phases, allow us to derive additional insights on the sources of these gains.
 The use of double buffering on the GPU side to overlap kernel processing with the device to host transfer in the merge phase is the largest source of gains (DtH in $\text{\sys}_\text{basic}$, which is replaced by processing time in the optimized \sys).
 %and, despite the device to device cost has a relatively larger cost for the smallest execution periods, the gains it enables largely outweigh the costs it imposes.
 On the CPU side, the ability to overlap transaction processing (noted \textit{non-blocking} in the figure) with the shipping of logs to the GPU has also a meaningful impact on reducing the blocking time, although not as strong as on the GPU side.

Finally, let us analyze the results reported in the right plot of Figure~\ref{fig:overhead_cmp}, which refers to the workload with 10\% of update transactions. In this scenario, which considers a less extreme (and arguably more realistic) application workload, the peak throughput of \sys converges to 4M tx/sec, which is very close to the peak throughput achievable by an idealized solution that achieves a performance equal to that of the two devices --- an additional evidence of the efficiency of the proposed design.

\subsection{Sensitivity to contention}
\label{sec:eval:cont}

We now consider the same workload as in the previous study, but inject with a given probability a conflicting access at random in the stream of writes generated by the CPU transactions.

\begin{figure}[tbp]
	\centering
	\includegraphics[width=0.6\columnwidth]{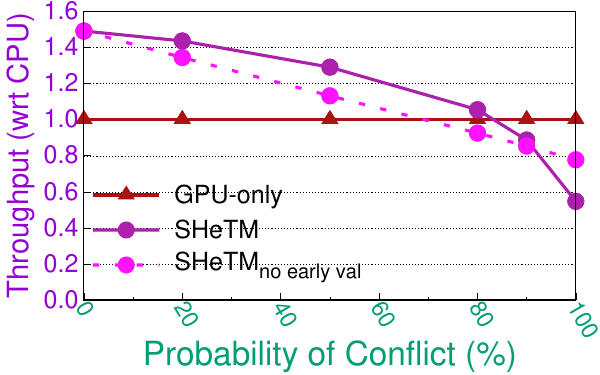}
	\vspace{-0.1cm}
	\caption{Sensitivity to inter-device contention.}
	\label{fig:inter_confl}
	\vspace{-0.5cm}
\end{figure}

We vary on the x-axis the inter-device conflict probability, fix the duration of the execution phase at 80 msecs and compare, in Figure~\ref{fig:inter_confl}, the performance of \sys with and without the early validation mechanism. On the y-axis we report the throughput normalized with respect to TSX (unistrumented) running solo and report, as reference, also the throughput achieved using PR-STM, running solo with double buffering.

The analysis of this plots reveals several  insights. The first observation is that \sys consistently outperforms both TSX and PRSTM for abort rates as high as 80\%. In medium contention, e.g., 50\% probability of contention, \sys continues to deliver a 30\% gain over the fastest individual device (CPU).
Even when operating at the extreme 100\% abort rate it incurs only a modest overhead (approx.~20\% if the early validation is disabled). Overall, these results confirm the robustness of \sys performance even in adverse scenarios.

Early validation  appears to be a powerful mechanism to mitigate overhead, especially in medium-high contention scenario (50\% and 80\% abort rate). Only beyond 80\% inter-device contention SHeTM presents overheads w.r.t. CPU and GPU. Such scenarios are arguably non-representative of the desired operational region of any TM system. In 100\% inter-device contention, early validation fails constantly, triggering the completion of the current execution phase and device transfer of the CPU logs. This is logically equivalent to operate with a much shorter execution phase, which, as seen in Figure~\ref{fig:overhead_cmp}, tends to induce longer blocking periods of the CPU.

\subsection{MemcachedGPU}
\label{sec:eval:memcached}

MemcachedGPU~\cite{Hetherington15} extends Memcached, a popular in-memory object caching system, in order to use GPUs to serve lookup (GET) and update (PUT) requests for cached objects.

Its original implementation presents an ad-hoc synchronization mechanism to manipulate the contents of the cache held in GPU's memory. Besides being non-trivial, PUT operations are executed single threaded and block other concurrent GETs. Furthermore, CPU and GPU work in a pipeline fashion, i.e., they do not access concurrently the same data at the same time. \sys ameliorates all these issues while maintaining the cache's state synchronized both on the CPU and GPU.

A fixed number of sets compose a cache. $\langle \textit{key}, \textit{value} \rangle$ pairs are hashed into the target set, which has 8 slots and may contain up to that many pairs (8-way associative). We use keys/values of 16B/32B, respectively. The LRU replacement policy evicts old pairs based on per-slot timestamp.

The CPU implementation is straightforward: a worker thread computes the target set and then issues a transaction to perform the GET/PUT logic. On the GPU, the target set is searched in parallel
(as in the original MemcachedGPU implementation~\cite{Hetherington15}) non-transactionally and a transaction is used to retrieve/set the value (for GETs/PUTs, respectively) and to update the LRU metadata of the corresponding slot.

The resulting implementation scheme introduces intra-device conflicts between any concurrent operation with the same input key, as each GET changes the per slot timestamp.
To reduce the likelihood of inter-device contention CPU and GPU use distinct timestamps, which implies that the LRU policy is local to CPU or GPU, i.e., the pair freshness is only affected by device-local transactions. The key advantage of this approach is that CPU GETs do not conflict with GPU GETs (and vice-versa).
%Note that both timestamps are globally visible via \sys, one could leverage this behaviour to non-transactionally observe the other device timestamp for better accuracy.

Concurrent inter-device PUT operations conflict if they
access the same set, which guarantees that it is safe to use local LRU policies --- as concurrent updates to the same set by different devices will trigger an abort and be rejected by \sys. To this end, a per set timestamp (common to both the CPU and CPU) is updated whenever a PUT operation is executed at either device. 

%\textcolor{red}{
%Concurrent inter-device PUT operations conflict if they access the same set. Given that, in the GPU, only the target slot is updated, an additional per set metadata is written in both the GPU and CPU to ensure that the set is not corrupted by inter-device accesses.}

Concurrent inter-device GETs and PUTs that access the same key conflict only if it is the CPU to issue the PUT. In fact, since \sys attempts to serialize the transactions executed by the CPU before the ones of the GPU, if the GPU issues a PUT it is perfectly fine for the CPU to ``miss'' this update.

In this experiment, we use a cache with 1,000,000 sets, the workload is composed of 99.9\% of GETs and the object popularity follows a Zipfian distribution with parameter $\alpha=0.5$ --- which is a reasonable value for evaluating caches~\cite{Breslau90}.

%In this experiment, we use a cache with 1000000 sets, which corresponds to a size of approx. 480MB. The sets are 8-way associative, and the size of the key is 16 bytes while that of the value is 32 bytes. We use LRU as replacement policy in case of eviction. The workload is composed by 99.9\% of GETs and the object popularity follows a Zipfian distribution with parameter $\alpha=0.5$ --- which represent typical  is a common distribution when settings for evaluating caches~\cite{Breslau90}.

We consider 4 different workloads. In the first workload (\textit{no-conflicts}), we balance the load (i.e., cache operations) input to the GPU and CPU by using the last bit of the key accessed by an operation. This guarantees that the input queues of the CPU and GPU can never contain operations that access a common set, thus, excluding the possibility of inter-device contention.

\begin{figure}[tbp]
	\centering
	\includegraphics[width=0.55\columnwidth]{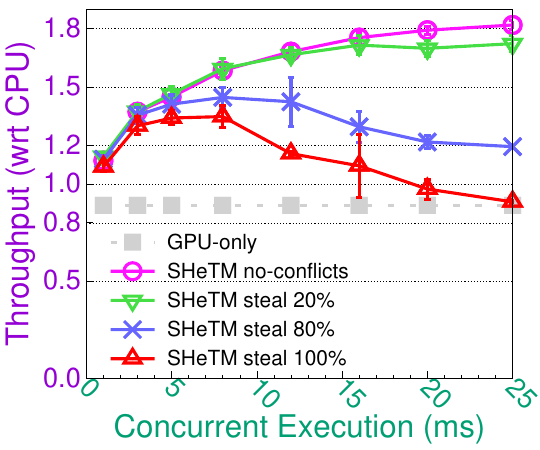}
	\vspace{-0.1cm}
	\includegraphics[width=0.43\columnwidth]{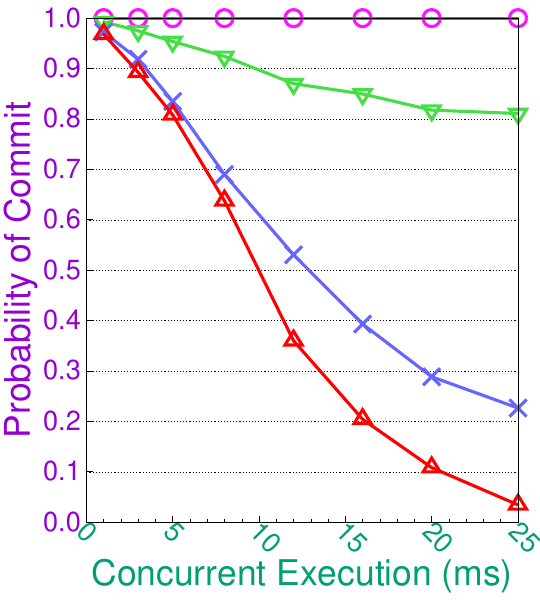}
	\caption{Throughput of HeTM for Memcached with possible conflicts.}
	\vspace{-0.5cm}
	\label{fig:memcd-3}
\end{figure}

We then evaluate load unbalanced scenarios, which
emulate application-level shifts of the popularity of accessed objects, namely a drop of the arrival rate of requests for $\text{GPU}_Q$ and a corresponding increase for $\text{CPU}_Q$. This causes the GPU to start stealing from the CPU queues, i.e., processing requests for objects that can be concurrently accessed by the CPU.
We consider three scenarios in which the GPU steals requests from the CPU queues with increasing probability, i.e., \textit{steal 20\%}, \textit{steal 80\%} and \textit{steal 100\%}. The last one emulates the extreme scenario where no device affinity is set to mitigate contention, so that both devices access the same keys.

Figure~\ref{fig:memcd-3} shows that \sys achieves almost indistinguishable performance in the \textit{no-conflicts} and the \textit{steal 20\%}, being in both cases less than 20\% away from the ideal solution and 80\% better than both GPU-only and CPU-only. 
We consider as an ideal solution the case that incurs no overhead and totals the equivalent normalized throughput of both CPU-only and GPU-only, which in this case is approx.~1.9.

The abort rate converges to that of the steal rate (resp. 0\%, 20\% 80\% and 100\%), since the number of potentially conflicting transactions executed at both devices increases when the duration of the execution round is increased. In fact, if its duration grows large enough, during a batch where one of the devices \textit{steals} transactions from the other, the probability of inter-device contention unavoidably converges to 1 for this workload. 
With the duration of approx.~25 msecs, execution rounds where GPU steals from the CPU are very likely to fail.

The gains remain significant even in case the GPU steals operations from the CPU queue with 80\% of probability (20\% to 40\% speed-up over CPU-only).
Finally, it is worth highlighting that even when the contention-avoidance dispatching mechanisms of \sys are not used and contention is high (steal 100\%), 
\sys achieves robust performance with speed-ups of up to approx.~30\% (10~msecs rounds), 
and throughput on par with CPU-only even when the inter-device conflict probability (right plot) converges to 1.

\section{Conclusions and Future Work}
\label{sec:concl}

This work introduced the abstraction of Heterogeneous  Transactional Memory (\abstraction). \hetm aims to facilitate programming of  heterogeneous  platforms, by abstracting the difficulties of data sharing across multiple physically separated units via the  illusion of a single transactional memory  shared among  CPUs and (discrete) GPU(s). 

%HeTM enables the possibility of accelerating irregular applications by offloading part of the computations, which would be executed concurrently by some CPU threads, to the GPU. Some concrete examples include K-means[3] (where CPU and GPU could process different data points and concurrently update the centroids) or the Parameter Server model for machine learning[4] (where the state of the model could be stored in STMR and be updated concurrently by worker threads running on both the CPU and GPU)

Besides introducing the abstract semantics and programming model of \abstraction, we presented an efficient, yet modular, implementation of the proposed abstraction, named Speculative \abstraction (\system). We demonstrated the efficiency of \system via an extensive quantitative study based both on synthetic benchmarks and on a popular object caching system.

This work opens a number of research questions related to defining alternative semantics and designs for the \hetm abstraction.
A specific question that we intend to investigate in the future is how to extend \sys to orchestrate the execution of multiple GPUs.

An other interesting question is how to leverage, e.g., static code analysis~\cite{Riegel08} or on-line scheduling techniques~\cite{Diegues15, Shrink} to automatically dispatch transactions to either CPU or GPU.

Finally, it would be interesting to evaluate \sys using a broader range of irregular applications, e.g., extending the recent work by Nelson et al.~\cite{Nelson19} that investigated the use of locking schemes to parallelize k-means on GPUs.

%\input{./sections/ack}
% \appendix
% TODO: Add Appendix later
%\input{./sections/2_1_demonstration}

%\clearpage % Submissions should be a maximum of ten (10) pages, not including references.
%\bibliographystyle{ACM-Reference-Format}
%\bibliographystyle{./template/pact19}
\bibliographystyle{IEEEtranS}
\bibliography{bibliography/library}

\end{document}